\documentclass{ptephy_v1}
\preprintnumber{XXXX-XXXX} 

\usepackage{indentfirst}

\newcommand{\apj}{Astrophys. J. }
\newcommand{\apjs}{Astrophys. J. Suppl. }

\begin{document}

\title{An approach to constrain models of  accreting neutron stars with the use of an equation of state}

\author{Akira Dohi}
\affil{Department of Physics, Kyushu University, Fukuoka 819-0395, Japan}
\author[1]{Masa-aki Hashimoto}
\author[1]{Rio Yamada}
\author[2]{Yasuhide Matsuo}
\affil{Nippo-cho, Minatokouhoku-ku, Yokohama, Kanagawa, 223-0057, Japan}
\author[]{Masayuki Y. Fujimoto}
\affil{Department of Physics, Hokkaido University, Sapporo 060-8810, Japan}

\begin{abstract}
We investigate X-ray bursts during the thermal evolution of an accreting neutron star which corresponds to the X-ray burster GS\ 1826-24. 
Physical quantities of the neutron star  are included using an equation of state below and above the nuclear matter density. We adopt an
equation of state and construct an approximate network that saves the computational time and calculates nuclear energy generation rates accompanying the abundance evolutions.
The mass and radius of the neutron star are got by solving the stellar evolution equations from the center to the surface
which involve necessary information such as the nuclear energy generation in accreting layers, heating from the crust, and neutrino emissions inside the stellar core.
We reproduce the light curve and recurrence time of the X-ray burst from GS 1826-24 within the standard deviation of 1$\sigma$
for the assumed accretion rate, metallicity, and
equation of state. It is concluded that the observed recurrence time is consistent with the theoretical model having
metallicity of the initial CNO elements $Z_{\rm CNO}$ = 0.01. We suggest that the nuclear reaction rates responsible for the $rp$-process
should  be examined in detail, because the rates may change the shape of the light curve and our conclusion.  
\end{abstract}

\subjectindex{E25,E32 Neutron stars, E34 Accretion}   
                     
\maketitle

\vspace{5pt}
\section{Introduction}

\begin{figure}[t]
\centering\includegraphics[width=0.6\linewidth]{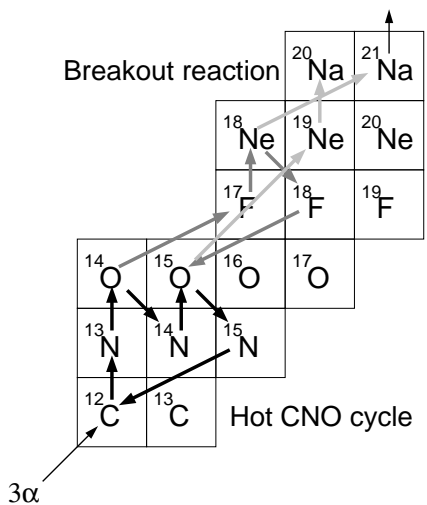}
\caption{Nuclear reaction sequence on a hydrogen accreting neutron star. After the Hot CNO cycle,
the breakout from the cycle begins through the two reactions $\rm^{14}O(\alpha,p)^{17}F$ and $\rm^{15}O(\alpha,\gamma)^{19}Ne$
accelerated by $\rm\alpha(2\alpha,\gamma)^{12}C$ reactions~\cite{Wallace1981}. Due to the increasing
temperature, the rapid proton capture process ($rp$-process) ensures by overcoming the Coulomb barrier.}
     \label{fig:1}
     \end{figure}

The type I X-ray burst was discovered from the Low Mass X-ray Binary in 1975~\cite{Grindlay1976} and 
112 bursters have been observed until now~\cite{Monash2019}. 
These observations provide firm ground to the arguments that
bursts are identified to be a phenomenon associated with shell flashes on the accreting neutron stars in close binary systems
 (e.g. \cite{Taam1982}).
The shell flashes are now believed to be initiated with thermonuclear runaway as shown in Fig.~\ref{fig:1}~\cite{Wallace1981}.
The X-ray burster GS 1826-24 has been observed in 1988 by X-ray astronomical satellite Ginga~\cite{Tanaka1989}.
Since an accretion rate for an X-ray burst to start generally varies even from the same burster, the shape of the light curve changes
significantly~\cite{Hashi2014}. 
However, the accretion rate ($dM/dt$) of GS 1826-24 has been known to be almost constant during a year and therefore the shape of the light curve is nearly invariant~\cite{Cornelisse2003}.
From the theory of the X-ray burst, it is convenient to make models of light curves under the constant
$dM/dt$~\cite{Fuji1981,Bildsten2000}.
As a consequence, it is 
believed to be appropriate to check the validity of models~\cite{Heger2007a,Lampe2016}.
Therefore, we can use the X-ray burster GS 1826-24 to compare numerical results of light curves calculated from the assumed $dM/dt$ with the observational data. In the present paper, we adopt the data of \textit{RXTE}  reported in Ref.~\cite{Galloway2017} to construct models.

In the meanwhile,
it was investigated to reproduce observational results by analyzing numerical results concerning X-ray bursts of
GS 1826-24~\cite{Heger2007a,Lampe2016,Meisel2018,Meisel2019}.
In Ref.~\cite{Heger2007a}, a large nuclear reaction network which includes 1300 nuclides was used to follow the detailed nucleosynthesis,
which is based on the numerical method by Woosley et al.~\cite{Woosley2004}. 
None the less, the numerical method does not include an equation of state at the high density above nuclear matter
which relates to light curves. This is because their calculations
have been limited to around the accretion layers above the crust, where the neutron star mass and radius are
selected as parameters to rebuild light curves. Furthermore,
artificial luminosity due to shallow and/or crustal heating
is given at the bottom of the calculated region which may locate above the outer crust of 
the neutron star~\cite{Zam2012}.
Generally speaking, it is difficult to constrain physical processes occurred inside the neutron stars as far as
only accretion layers responsible for bursts are taken account of~\cite{Fujimoto1984}. In this connection, the ignition properties have been
investigated from the point of  ``mixed bursts''~\cite{Nara2003}, where impact on the burst ignition related to stable and/or
unstable nuclear burnings has been examined with the use of a linear stability analysis; The discrepancies between the results and other
multi-zone calculations may be ascribed to the strength of the Hot CNO cycle~\cite{Nara2006}. Since many reaction rates responsible
for the $rp$-process are still very uncertain, the suggested problem remains unresolved. On the other hand, effects of the inner core
below the neutron star crust are not clear due to the uncertainty of nuclear physics at the high density, where an equation of state (EoS)
is still very uncertain. For example, cooling of neutron stars~\cite{Dohi2019,Lim2017} or supernova explosions~\cite{Oertel2017} have been studied
taking account of some EoSs having significant uncertainty.

In the present study, to perform the calculation of X-ray bursts, 
the general relativistic equations for the stellar structure and evolution are solved exactly from the center
to the outermost layer, where we can study thermal interactions between the accreted layers and the core of the neutron star~\cite{Fujimoto1984}. Besides,
we develop an approximate network that includes 88 nuclides; it reproduces the nuclear energy generation and rapid proton capture process ($rp$-process) during the burst. By including necessary physical inputs such as an equation of state, nuclear burning, opacities,
and neutrino losses, we propose some models with the mass and radius of the neutron star responsible for the X-ray burster GS 1826-24.
In section 2, basic equations of neutron star evolution are given with physical inputs. Our approximate network to follow the nuclear processes is explained in section 3.
The results of the calculations are compared with the observations using the statistical analysis in section 4. We discuss our results 
toward further understanding of X-ray bursts in section 5.

\noindent
\section{Basic equations and physical inputs for accreting neutron stars}

We have performed numerical calculations of the thermal evolution of neutron stars in hydrostatic equilibrium by using the spherical symmetric stellar evolutionary code~\cite{Fujimoto1984}, which includes full general relativistic effects  formulated by Throne~\cite{Thorne1977}.
The basic simultaneous differential equations are written as follows:
\begin{eqnarray}
  \frac{\partial M_{tr}}{\partial r} \hspace*{-1mm}& = &\hspace*{-1mm} 4\pi r^{2} \rho~, \label{eq:1} \\
  \frac{\partial P}{\partial r}\hspace*{-1mm} & = &\hspace*{-1mm} -\frac{GM_{tr}\rho}{r^{2}}
      \left(1+\frac{P}{\rho c^{2}}\right)
      \left(1+\frac{4\pi r^{3}P}{M_{tr}c^{2}}\right) \left(1-\frac{2GM_{tr}}{c^{2}r}\right)^{-1}~, \label{eq:2} \\
  \frac{\partial (L_{r}e^{2\phi/c^{2}})}{\partial M_{r}}\hspace*{-1mm} & =\hspace*{-1mm} &
      e^{2\phi/c^{2}}\left(\varepsilon_{\rm n}+\varepsilon_{\rm g}-\varepsilon_{\nu}
      \right)~, \label{eq:3} \\
  \frac{\partial \ln T}{\partial \ln P}\hspace*{-1mm} & =\hspace*{-1mm} & {\rm min}(\nabla_{\rm rad}, \nabla_{\rm ad})~, \label{eq:4} \\
  \frac{\partial M_{tr}}{\partial M_{r}}\hspace*{-1mm} & =\hspace*{-1mm} & \frac{\rho}{\rho_0}
  \left(1-\frac{2GM_{tr}}{c^{2}r}\right)^{1/2}~, \label{eq:5}\\
  \frac{\partial \phi}{\partial M_{tr}}\hspace*{-1mm} & =\hspace*{-1mm} & \frac{G(M_{tr}+4\pi r^{3}P/c^{2})}
    {4\pi r^{4}\rho}\left(1-\frac{2GM_{tr}}{c^{2}r}\right)^{-1}. \label{eq:6} %
\end{eqnarray}
Here, $M_{tr}$ and $M_r$ in a radius $r$ are gravitational and rest masses;
$\rho$ and $\rho_0$ denote the total mass-energy and rest mass densities;
$P$ and $T$ are the pressure and local temperature;
$\varepsilon_{\rm n}$ and $\varepsilon_{\rm g}$ are the energy generation rates by nuclear burning and gravitational energy release, respectively. 
Besides, $\varepsilon_\nu$  represents the energy loss rate by neutrino emission;
$\nabla_{\rm rad}$ and $\nabla_{\rm ad}$ are the radiative and adiabatic gradients;
$\phi$ is the gravitational potential in unit mass.

We adopt EoS by Lattimer and Swesty~\cite{Lattimer1991} with the incompressibility $K$ of 180 or 220 MeV in the inner layers~($\rho\geq10^{12.8} \rm~g~cm^{-3}$) and connect it to EoS in Refs. \cite{Fujimoto1984,Baym1971} for the outer layers
($\rho<10^{12.8} \rm~g~cm^{-3}$). It is noted that the EoS we have adopted is still applied to study the neutron star properties above the
nuclear density~\cite{Lim2017,Oertel2017}, because it may be consistent with both experimental and astrophysical constraints compared to
other sophisticated EoS~\cite{Dohi2019}.
Opacities appropriate for the neutron star evolution are the same as in Ref.~\cite{Matsuo2018}.
The neutrino emission process is taken into account as a slow cooling process: electron-positron pair, photo, and plasmon processes~\cite{Festa1969,Friman1979}; bremsstrahlung process; modified Urca (MURCA) process.
Dominant processes are MURCA~\cite{Chiu1964} and bremsstrahlung~\cite{Festa1969}.
The corresponding energy loss rates are summarized in Ref.~\cite{Friman1979}:
Although these neutrino emission rates have been studied, there remains the uncertainty of a factor of ten because of  insufficient understanding of the symmetry energy and nucleon effective mass in the dense matter \cite{Yin2017}.
Moreover, we do not include a strong process such as pion condensation.
While pion condensation accompanies strong neutrino emissions, the effects of the super-fluidity may reduce the neutrino emissions below
the critical temperature $T_{\rm cr}$ in proportion to $\exp(-a T_{\rm cr}/T)$ with a constant $a$.
Since the present purpose is to show our approach to constrain the properties coming from the neutron star core 
using X-ray burst observations for high accretion rates $dM/dt > 10^{-9} M_\odot~\rm yr^{-1}$, we neglect the strong neutrino emission process~\cite{Fujimoto1984}.

We construct initial models of a neutron star accreted at  the constant rates 
$dM/dt =(2 - 4) \times 10^{-9} M_\odot~\rm yr^{-1}$
with mass fractions, $X(^{1}{\rm H}) = 0.73$, $X(^{4}{\rm He}) = 0.25$, $X(^{14}{\rm O}) = 0.007$, and  $X(^{15}{\rm O}) = 0.013$,
where $X(\rm O^{14})$ and $X(\rm O^{15})$ are assumed to be equilibrium values of Hot~CNO cycle (see Fig.~\ref{fig:1}).
The initial model  is in a steady-state, in which the non-homologous part of the gravitational energy release can be
neglected~\cite{Fujimoto1984}, where nuclear burning is temporally switched off.
As a result, we can get initial models for $\dot{M}_{-9} =$ 2, 2.5, 3, and 4, where $\dot{M}_{-9}$ is the accretion rate in units of
$10^{-9}~M_\odot~\rm yr^{-1}$. 
Corresponding to the mass-radius relation against $K$,  the gravitational mass and radius are obtained 
as follows:  ($M/M_\odot$, $R(\rm km)$) = (1.57, 11.9) for $K=180$~MeV,  (1.58, 12.6) and (2.0, 11.3) for $K=220$~MeV.
The mass of $2~M_{\odot}$ may be rather high for the low mass X-ray binary; we adopt this model for reference
as shown in Table~\ref{tab:result}.

		\begin{table}
			\caption{New approximate network (APRX3) which includes 88 nuclides improved after Ref.~\cite{Wallace1981,Hanawa1983}.}
			\label{APRX3}
			\begin{center}
				\begin{tabular}{cccccc} \vspace{-0.45cm} \\ \hline\hline
					Nuclides & $A$ & Nuclides & $A$ & Nuclides & $A$ \\
					\hline
					\multicolumn{1}{l}{H} & \multicolumn{1}{r|}{1} &
					\multicolumn{1}{l}{Cr} & \multicolumn{1}{r|}{46} &
					\multicolumn{1}{l}{Tc} & \multicolumn{1}{r}{92} \\
					\multicolumn{1}{l}{He} & \multicolumn{1}{r|}{4} &
					\multicolumn{1}{l}{Fe} & \multicolumn{1}{r|}{48, 50} &
					\multicolumn{1}{l}{Ru} & \multicolumn{1}{r}{88, 90, 92} \\
					\multicolumn{1}{l}{C} & \multicolumn{1}{r|}{12} &
					\multicolumn{1}{l}{Ni} & \multicolumn{1}{r|}{54, 56, 60} &
					\multicolumn{1}{l}{Rh} & \multicolumn{1}{r}{96} \\
					\multicolumn{1}{l}{O} & \multicolumn{1}{r|}{14 -- 16} &
					\multicolumn{1}{l}{Zn} & \multicolumn{1}{r|}{60, 64} &
					\multicolumn{1}{l}{Pd} & \multicolumn{1}{r}{92, 94, 96, 98} \\
					\multicolumn{1}{l}{Ne} & \multicolumn{1}{r|}{18} &
					\multicolumn{1}{l}{Ge} & \multicolumn{1}{r|}{62 -- 64, 68} &
					\multicolumn{1}{l}{Ag} & \multicolumn{1}{r}{97 -- 98, 102} \\
					\multicolumn{1}{l}{Mg} & \multicolumn{1}{r|}{21 -- 22} &
					\multicolumn{1}{l}{Se} & \multicolumn{1}{r|}{68, 72} &
					\multicolumn{1}{l}{Cd} & \multicolumn{1}{r}{96, 98, 100,} \\
					\multicolumn{1}{l}{Si} & \multicolumn{1}{r|}{24 -- 25} &
					\multicolumn{1}{l}{Kr} & \multicolumn{1}{r|}{72, 76} &
					\multicolumn{1}{l}{} &  \multicolumn{1}{r}{102 -- 106} \\
					\multicolumn{1}{l}{S} & \multicolumn{1}{r|}{28 -- 30} &
					\multicolumn{1}{l}{Sr} & \multicolumn{1}{r|}{76, 80} &
					\multicolumn{1}{l}{In} & \multicolumn{1}{r}{99, 102 -- 107, 109} \\
					\multicolumn{1}{l}{Ar} & \multicolumn{1}{r|}{33 -- 34} &
					\multicolumn{1}{l}{Zr} & \multicolumn{1}{r|}{80, 84} &
					\multicolumn{1}{l}{Sn} & \multicolumn{1}{r}{100 -- 109, 112} \\
					\multicolumn{1}{l}{Ca} & \multicolumn{1}{r|}{37 -- 40} &
					\multicolumn{1}{l}{Nb} & \multicolumn{1}{r|}{88} &
					\multicolumn{1}{l}{Sb} & \multicolumn{1}{r}{106 -- 108} \\
					\multicolumn{1}{l}{Ti} & \multicolumn{1}{r|}{42} &
					\multicolumn{1}{l}{Mo} & \multicolumn{1}{r|}{84} & 
					\multicolumn{1}{l}{Te} & \multicolumn{1}{r}{107 -- 109}\\ \hline
				\end{tabular}
			\end{center}
		\end{table}

\vspace{5pt}
\noindent
\section{Approximate network}

We have constructed an approximate network that nearly reproduces the nuclear energy generation rates and nuclear abundances related to X-ray bursts~\cite{Wallace1981,Hanawa1983}.
The small network was developed  first by Wallace and Woosley~\cite{Wallace1981} with 10 nuclei up to $\rm^{56}Ni$
and then by Hanawa et al.~\cite{Hanawa1983} with the inclusion of 16 nuclei up to $^{68}$Se.  The construction of these approximate networks
is based on the calculations of the nucleosynthesis by large networks. We have made a revised version of the approximate network
(hereafter APRX3) which includes 88 nuclei as shown in Table~\ref{APRX3}. To construct APRX3, we used a large network~\cite{Koike1999}, where the updated nuclear
reaction rates have been adopted from Ref.~\cite{Cyburt2010} except for some data: Two reactions 
$\rm^{64} Ge(p,\gamma)^{65}As$ and $\rm^{65}As(p,\gamma)^{66}Se$ and their reverse rates are taken from Ref.~\cite{Lam2016}. 
The $\beta^+$ rates are from Refs.~\cite{Hashimoto1985,Oda1994,Langanke2000,Nabi1999}.
APRX3 is built to reproduce the nuclear reaction paths examined by
Schatz~\cite{Schatz2006}. In particular, it is pointed out that the $rp$-process proceeds to the formation of nuclei of $Z<50$ and
the nucleosynthesis is closed down at the Sn-Sb-Te cycle:
			\begin{equation}\label{SnSbTe}
\rm
			^{103}{Sn}(\beta)^{103}{In}(p,\gamma)^{104}Sn(\beta)^{104}{In}(p,\gamma)^{105}{Sn}(p,\gamma)
			^{106}{Sb}(p,\gamma)^{107}{Te}(\gamma,\alpha)^{103}{Sn}. \nonumber
			\end{equation}

\begin{figure}[tp]
\begin{center}
    \begin{tabular}{lcr}
    \begin{minipage}[t]{0.49\hsize} 
       \mbox{\raisebox{-1.7mm}{
       \includegraphics[width=1.0\linewidth]{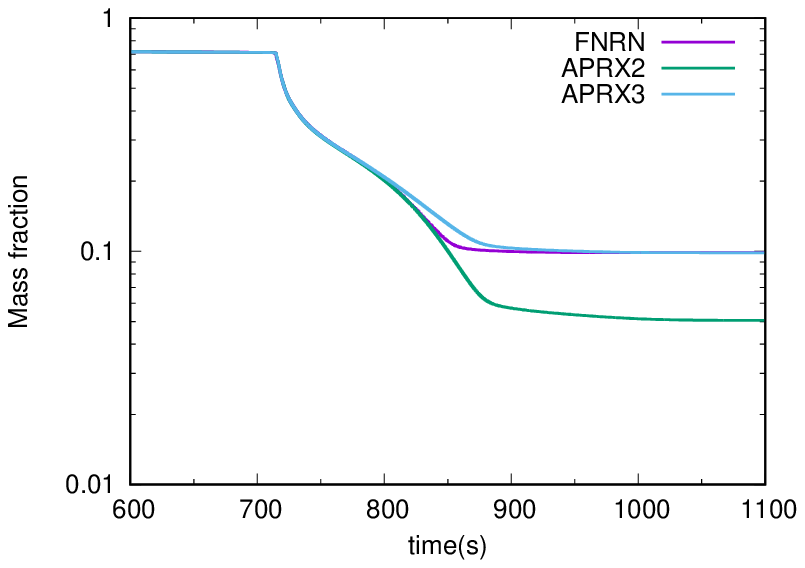}}}
\caption{Changes in $X(^{1}{\rm H})$ under the pressure of $\log~P=22.8$ during 600--1100 s.}
     \label{fig:hydrogen}
    \end{minipage}
    &\,&
    \begin{minipage}[t]{0.49\hsize}
        \mbox{
        \includegraphics[width=1.0\linewidth]{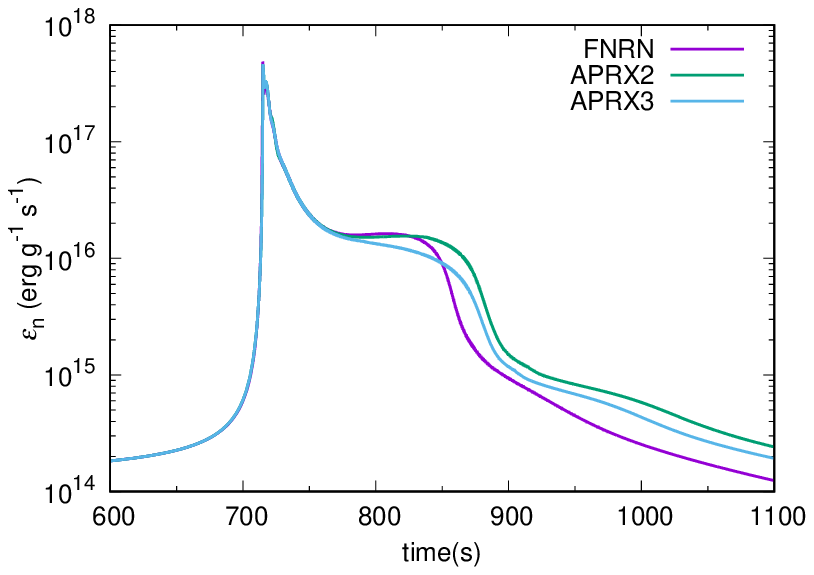}}
        \caption{Same as Fig.~\ref{fig:hydrogen} except for the nuclear energy generation rate.}
        \label{fig:nucene}
    \end{minipage}
    \end{tabular}
\end{center}
\end{figure}

Our large nuclear reaction network contains 897 nuclei up to $^{130}\rm{Sm}$ (hereafter FNRN) and is
adequate to study  X-ray bursts~\cite{Koike1999}. To simulate bursts we assume one zone model of the ignition pressure
to be $\log P=22.8$ for the neutron star with $M=1.4M_{\odot}$ and $R=10~\rm km$~\cite{Hanawa1983,Koike1999}. The initial mass fractions are the same as described in the previous section.
The networks reproduce changes in hydrogen abundance and nuclear energy generation rate
as shown in Figs.~\ref{fig:hydrogen} and \ref{fig:nucene}, respectively, where
APRX2 is another approximate network that contains 61nuclei~\cite{Matsuo2017}. It can be seen that $X\rm^1(H)$ 
calculated by APRX2 deviates
significantly from that of FNRN at $t>850$~s.  APRX3 can reproduce both $X(\rm^1H)$ and $\varepsilon_{\rm n}$ of FNRN with enough
accuracy. Therefore, we adopt APRX3 in the present study.
In the above calculations, the computational time of FNRN is 6.3 times longer than that of APRX3. We note that our parallel computer machine has the power of  fourteen cores per cpu.

\vspace{5pt}
\noindent
\section{Comparison with observations}

The observed luminosity $ L_{\rm b}^{\rm obs} $ can be expressed in terms of the observed flux $F_{\rm b}^{\rm obs}$ during a burst and the
distance $d$ to GS 1826-24 as follows:
		\begin{equation}\label{4.1}
			L_{\rm b}^{\rm obs} = 4 \pi d^2 \xi_{\rm b} F_{\rm b}^{\rm obs},
		\end{equation}
where $\xi_{\rm b}$ is a factor of anisotropy for the burst; if the radiation is isotropic, $\xi_{\rm b}=1$. It is noted that the calculated peak
luminosity $L_{\rm pk}$ divided by $(1+z)^2$ corresponds to $L_{\rm b}^{\rm obs}$.
On the other hand, $d^2\xi_{\rm b}$ is not obtained from observations. Therefore, $L_{\rm b}^{\rm obs}$ cannot be compared to
the calculated value of  $L_{\rm b}^{\rm thr}$.
In our case, we define $\chi^2$ at a given trial value of $d\xi_{\rm b}^{1/2}$,
		\begin{equation}\label{key}
			\chi^2 \equiv \sum_{i}^{n_{\rm{obs}}} \frac{(L_{{\rm b}, i}^{\rm obs}(d \xi_{\rm b}^{1/2}, t_{\rm pk}) - L_{\rm b}^{\rm thr})^2}{\sigma_{i,\rm{obs}}^2+\sigma_{i,\rm{thr}}^2}.     
		\end{equation}
Here, $t_{\rm pk} $ is the time from the beginning to the peak of the luminosity, $ n_{\rm{obs}} $ is the total number of observations 
consisting of the light curve ($ n_{\rm{obs}}=10$)~\cite{Galloway2017},
$L_{{\rm b}, i}^{\rm obs}(d \xi_{\rm b}^{1/2},t_{\rm pk})$ is the luminosity under the assumption of two assumed parameters of
$d \xi_{\rm b}^{1/2}$ and $t_{\rm pk}$, and $L_{\rm b}^{\rm thr}$ is the theoretical luminosity calculated from our models.
Two values of the standard deviations, $ \sigma_{i,\rm obs }$ and $\sigma_{i,\rm thr }$ are those of observation and calculation, respectively.
We note that $\sigma_{i,\rm thr }$ is taken to be the deviation from the average luminosity over the total number of the successive 
luminosities $n_{\rm burst}$.
The minimum quantity of $\chi^2$ is determined by changing both $d\xi_b^{1/2}$ and $t_{\rm pk}$.

\begin{figure}[t]
\begin{minipage}{0.49\textwidth}
\centering
\includegraphics[width=\linewidth]{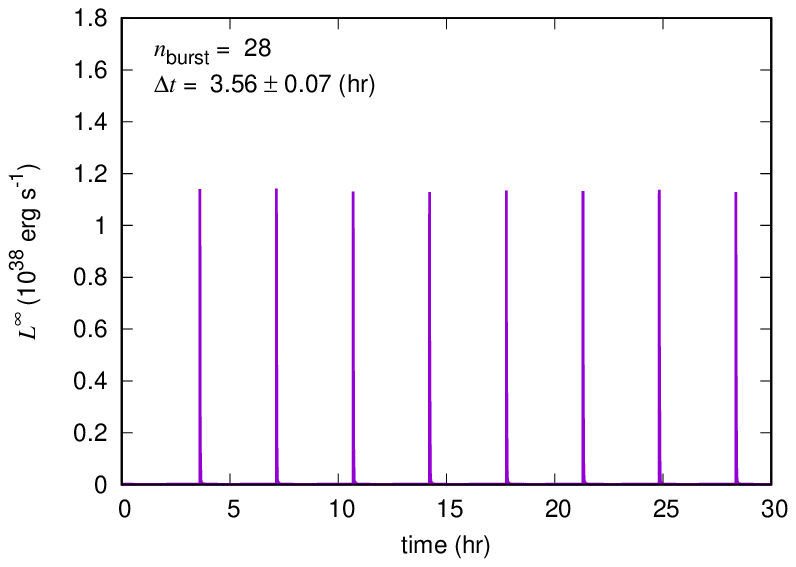}
\end{minipage}
\begin{minipage}{0.49\textwidth}
\centering
\includegraphics[width=\linewidth]{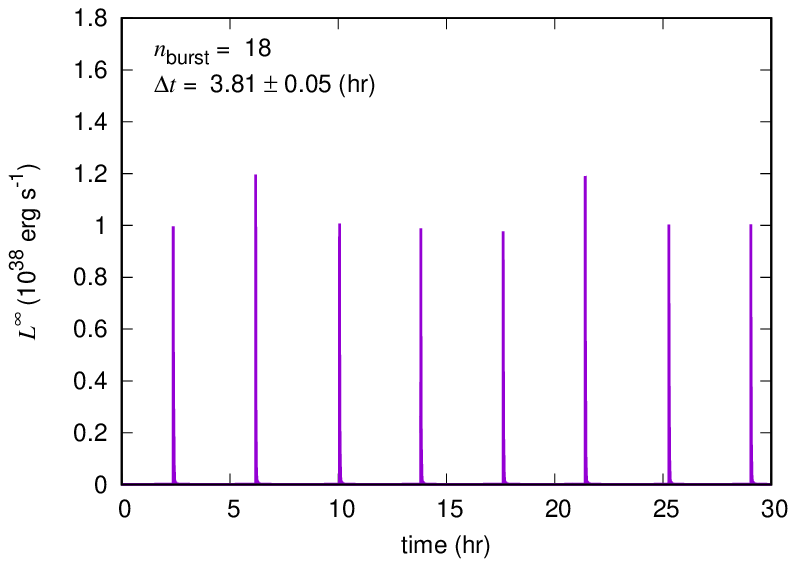}
\end{minipage}
\caption{Light curves during $0$ -- $30~{\rm hr}$ 
with $\dot{M}_{-9} = 2.0$ and $M = 1.58~M_{\odot}$. Left panel: $Z_{\rm CNO} = 0.01$. Right panel: $Z_{\rm CNO} = 0.005$.
We insert $\Delta t$ and numbers of bursts $n_{\rm burst}$ used for the present analysis.}
\label{z0015}
\end{figure}

\begin{figure}
\begin{minipage}{0.49\textwidth}
\centering
\includegraphics[width=\linewidth]{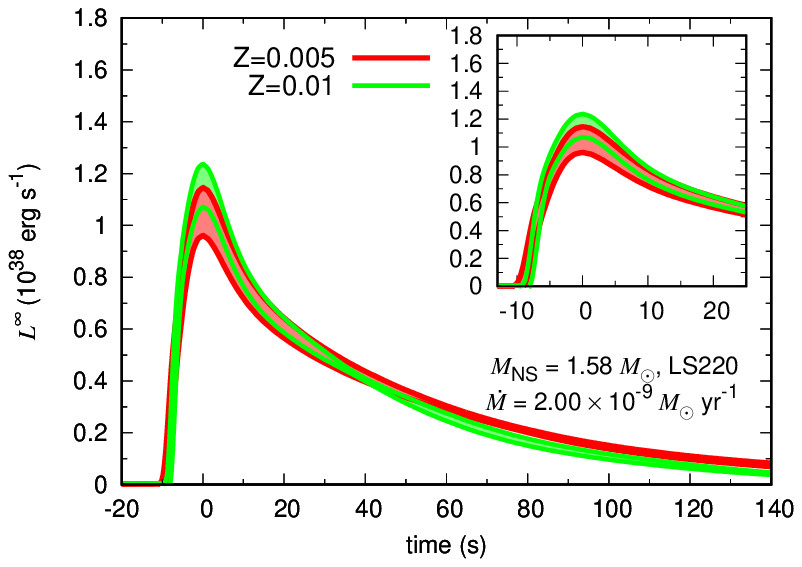}
\end{minipage}
\begin{minipage}{0.49\textwidth}
\centering
\includegraphics[width=\linewidth]{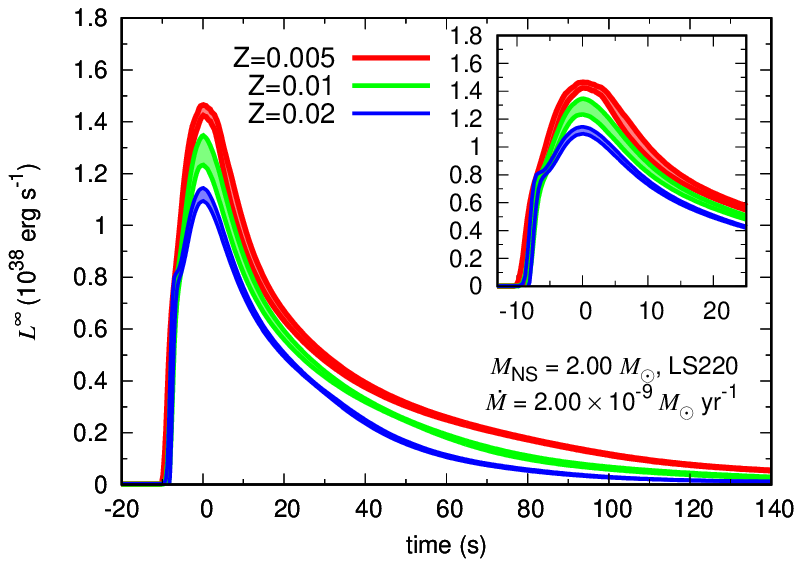}
\end{minipage}
\caption{Light curves with $\dot{M}_{-9} = 2.0$ with 1 $\sigma$ regions. We set the peak at $t = 0~{\rm s}$. Red color shows the region of $Z_{\rm CNO} = 0.005$, green $Z_{\rm CNO} = 0.01$, and blue $Z_{\rm CNO} = 0.02$, respectively. Left panel corresponds to  $M = 1.58~M_{\odot}$ while right $M = 2.00~M_{\odot}$. The model with $Z_{\rm CNO} = 0.02$ and $M = 1.58~M_{\odot}$ is not drawn 
because  $L_{\rm pk}$ is higher than the Eddington luminosity.}
\label{z005to2}
\end{figure}

In Table~\ref{tab:result}, the results of calculations are summarized for 27 models, where we can see the effects of EoS, 
the accretion rate, and the metallicity on the light curve.
 Models LS180 and LS220 adopt EoS~\cite{Lattimer1991} with the respective incompressibility $K$ of 180 and 220 MeV;
 the accretion rate $\dot{M}_{-9}$ =2, 2.5, 3, and 4. The metallicity of CNO elements is selected to be $Z_{\rm CNO}$ = 0.005, 0.01, and 0.02. 
Eight light curves are shown for models L2n20Z1 and L2n20Z05 in Fig.~\ref{z0015}.
We note that the recurrence time  $\Delta t$ depends on $\dot M$, $M$, and $Z_{\rm CNO}$.
In Fig.~\ref{z005to2}, it is seen that the light curves depend significantly on the metallicity, which constrains the model;
Both the height of the peak luminosity and the shape of the light curve are different in obedience to $Z_{\rm CNO}$.
We note that it is difficult to fit both the height of the light curve and the recurrence time in case of 
$M=2~M_{\odot}$.  

For comparison, we show the results in the table with the use of $K=180$~MeV whose value is incapable of reproducing the experimental one of 
$K$~\cite{Hebe2013,Garg2018},
where the calculated luminosity is lower than that of the observed one.

\begin{figure}[htbp]
\begin{minipage}{0.49\textwidth}
\centering
\includegraphics[width=\linewidth]{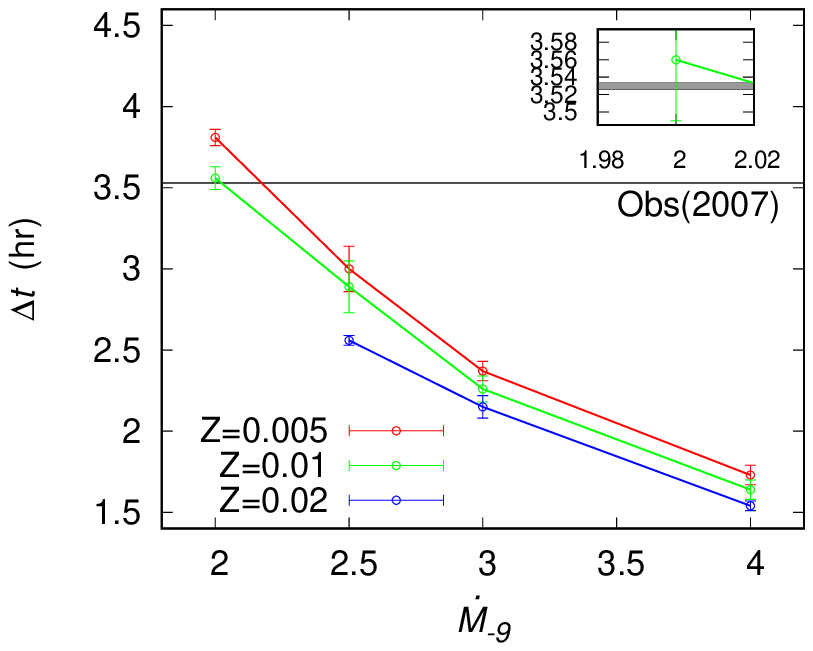}
\end{minipage}
\begin{minipage}{0.49\textwidth}
\centering
\includegraphics[width=\linewidth]{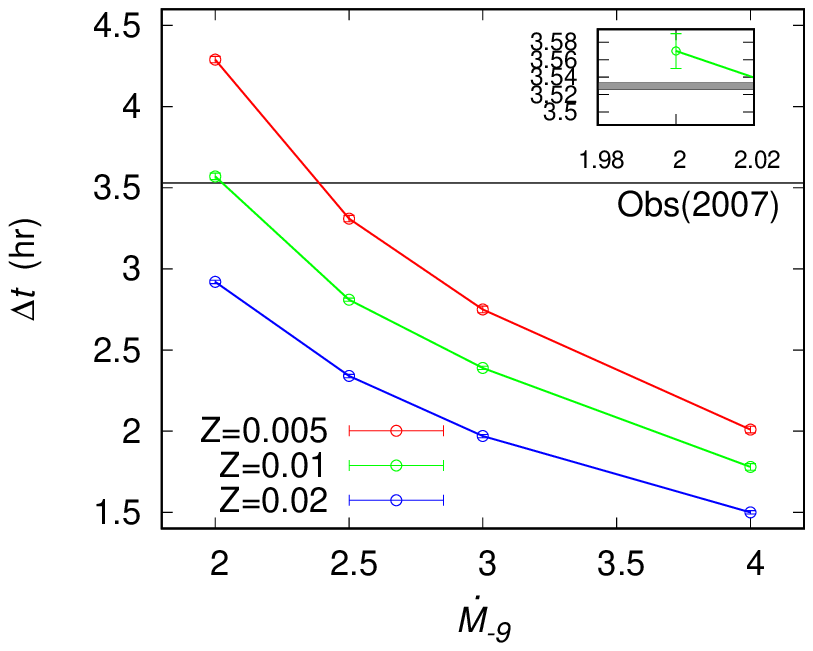}
\end{minipage}
\caption{Recurrence time $\Delta t$ v.s. accretion rate $\dot{M}_{-9}$. The error bars are 1$\sigma$ regions of $\Delta t$.
From the latest observational value of GS1826-24 in 2007, we adopt $\Delta t = 3.530\pm 0.004~{\rm hr}$ as indicated by s black line. The insets in these upper-right corners indicate magnification of the region around $\Delta t=3.53$~hr and $\dot{M}_{-9} = 2.0$. 
Left panel corresponds to  $M = 1.58~M_{\odot}$ while right $M = 2.00~M_{\odot}$.
}
\label{deltat}
\end{figure}

Figure~\ref{deltat} designates the recurrence time between bursts against the accretion rate for the calculated models.
It depends significantly on the metallicity. As shown in Table~\ref{tab:result} we obtained the recurrence time
$\Delta t =3.56\pm0.07$~hr for L2n20Z1
which is in the range of the observed value of $\Delta t =3.536\pm0.04$~hr~\cite{Galloway2017}. 
In addition to the recurrence time, it is clear that the overall shape of the light curve for the lower-mass model (L2n20z1) is much better than
the higher mass model such as L2n20Z1 as shown in Fig.~\ref{fig:bestfit}. 

Since other models cannot fit the peak value of the light curve observation, among models investigated in the present
paper, {\it the best fit model} which reproduces the observation of 
the light curve in 2007~\cite{Galloway2017} becomes L2n20Z1, where
we can see the degree of agreement between calculations with $1\sigma$ band and observations.
We remark that while the observational average of the light curve was obtained from 10 bursts ($n_{\rm burst}=10$), we select 21 successive
bursts ($n_{\rm burst}=21$) to get average luminosity. We may eliminate numerical ambiguity by including more bursts compared to actual observations. It should be noted that
the model of L2n20z1 agrees with the light curve observation as well as the result of Heger et al. (see their figure 1)~\cite{Heger2007a}, where they used 30 bursts having the large nuclear reaction
network with 1300 nuclides included. 


\begin{table}[htbp]
		\caption{Physical quantities for 27 models.  We adopt the observational data in 2007~\cite{Galloway2017}.
Numerical value inside the bracket indicates the standard deviation 1$\sigma$.}
		\label{tab:result}
		\begin{center}
			\scalebox{0.95}{
			\begin{tabular}{ccccccccccc} \vspace{-0.45cm} \\ \hline\hline
				MODEL & EoS & $ M $ & $ R $ & $ \dot{M}_{-9}$ & $Z_{\rm{CNO}}$ & $ L_{\rm{pk}} $ & $ \Delta t $ & $d\sqrt{\xi_b}$(2007) & $\chi^2_{\nu,\mathrm{min}}(2007)$\\
				&&$ [ M_\odot ] $&$ [\rm{km}]  $&  &  &  $ [10^{38}\rm{erg\ s^{-1}}] $& [hr] & [kpc] &  \\
				\hline
				%
%
L1n30Z05 &LS180 & 1.57 & 11.9 & 3.0 & 0.005  & 0.92 (0.04) & 2.23 (0.05)  & 5.69 & 4.75 \\%
				L1n30Z1 &LS180 & 1.57 & 11.9 & 3.0 & 0.01 & 0.95 (0.04) & 2.08 (0.03) & 5.59 & 1.42  \\%
L1n30Z2 &LS180 & 1.57 & 11.9 & 3.0 & 0.02 & 1.07 (0.11) & 1.91 (0.06)  & 5.51 & 3.07  \\%
L2n20Z05 &LS220 & 1.58 & 12.6 & 2.0 & 0.005  & 1.05 (0.09) & 3.81 (0.05)  & 6.13 & 2.04  \\%
L2n20Z1 &LS220 & 1.58 & 12.6 & 2.0 & 0.01  & 1.16 (0.08) & 3.56 (0.07) & 6.29 & 1.48  \\
L2n20Z2 &LS220 & 1.58 & 12.6 & 2.0 & 0.02  & - & -  & - & - \\%
L2n25Z05 &LS220 & 1.58 & 12.6 & 2.5 & 0.005 & 1.08 (0.13) & 3.00 (0.14)   & 6.09 & 0.83 \\%
L2n25Z1 &LS220 & 1.58 & 12.6 & 2.5 & 0.01 & 1.34 (0.25) & 2.89 (0.16) & 6.10 & 1.23    \\%
L2n25Z2 &LS220 & 1.58 & 12.6 & 2.5 & 0.02 & 1.20 (0.02) & 2.56 (0.03)  & 5.54 & 2.30   \\%
L2n30Z05 &LS220 & 1.58 & 12.6 & 3.0 & 0.005 & 0.98 (0.06) & 2.37 (0.06)  & 5.92 & 4.18   \\%
L2n30Z1 &LS220 & 1.58 & 12.6 & 3.0 & 0.01 & 1.06 (0.11) & 2.26 (0.08) & 5.93 & 0.35    \\%
L2n30Z2 &LS220 & 1.58 & 12.6 & 3.0 & 0.02 & 1.22 (0.16) & 2.15 (0.07) & 5.88 & 1.17   \\%
L2n40Z05 &LS220 & 1.58 & 12.6 & 4.0 & 0.005 & 0.89 (0.06) & 1.73 (0.06) & 5.82 & 6.78\\%
L2n40Z1 &LS220 & 1.58 & 12.6 & 4.0 & 0.01 & 0.94 (0.08) & 1.64 (0.06) & 5.71 & 1.68\\%
L2n40Z2 &LS220 &  1.58 & 12.6 & 4.0 & 0.02  & 1.02 (0.05) & 1.54 (0.03) & 5.60 & 2.36  \\%
%
%
L2h20Z05 &LS220 & 2.00 & 11.3 & 2.0 & 0.005  & 1.44 (0.02) & 4.29 (0.02) & 6.43 & 5.41  \\%
L2h20Z1 &LS220 & 2.00 & 11.3 & 2.0 & 0.01 & 1.29  (0.06) & 3.57 (0.02)  & 5.87 & 10.15  \\
L2h20Z2 &LS220 & 2.00 & 11.3 & 2.0 & 0.02 & 1.13 (0.03) & 2.92  (0.01)  & 5.57 & 30.02 \\%
L2h25Z05 &LS220 & 2.00 & 11.3 & 2.5 & 0.005  & 1.27 (0.03) & 3.31 (0.02)   & 6.10 & 4.87 \\%
L2h25Z1 &LS220 & 2.00 & 11.3 & 2.5 & 0.01 & 1.15 (0.02) & 2.81 (0.01) & 5.77 & 6.53  \\%
L2h25Z2 &LS220 & 2.00 & 11.3 & 2.5 & 0.02  & 1.13 (0.03) & 2.34 (0.01) & 5.50 & 22.36 \\%
L2h30Z05 &LS220 & 2.00 & 11.3 & 3.0 & 0.005  & 1.26 (0.04) & 2.75 (0.02)  & 6.06 & 3.48 \\%
L2h30Z1 &LS220 & 2.00 & 11.3 & 3.0 & 0.01  & 1.08 (0.04) & 2.39 (0.01)  & 5.68 & 2.68 \\%
L2h30Z2 &LS220 & 2.00 & 11.3 & 3.0 & 0.02  & 1.08 (0.04) & 1.97 (0.01)  &  5.35 & 7.66 \\%
L2h40Z05 &LS220 & 2.00 & 11.3 & 4.0 & 0.005 & 1.16 (0.01) & 2.01 (0.02) & 5.91 & 2.49\\%
L2h40Z1 &LS220 & 2.00 & 11.3 & 4.0 & 0.01 & 1.11 (0.05) & 1.78 (0.02) & 5.68 & 2.68 \\%
L2h40Z2 &LS220 & 2.00 & 11.3 & 4.0 & 0.02  & 1.03 (0.04) & 1.50 (0.01) &  5.35 & 7.66 \\%
				\hline
			\end{tabular}}
		\end{center}
\end{table}

\vspace{9pt}

\newpage
\noindent

\section{Discussion}

\begin{figure}[htbp]

\begin{minipage}{0.49\textwidth}
\centering
\includegraphics[width=\linewidth]{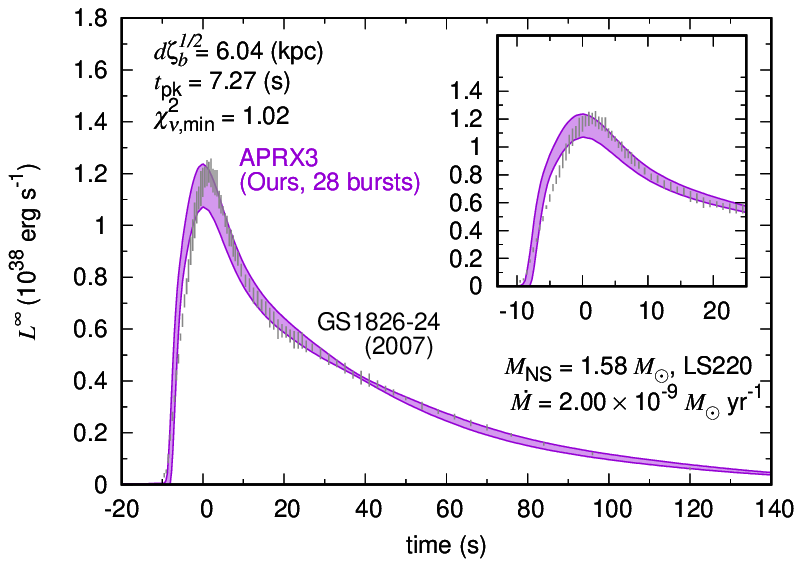}
\end{minipage}
\begin{minipage}{0.49\textwidth}
\centering
\includegraphics[width=\linewidth]{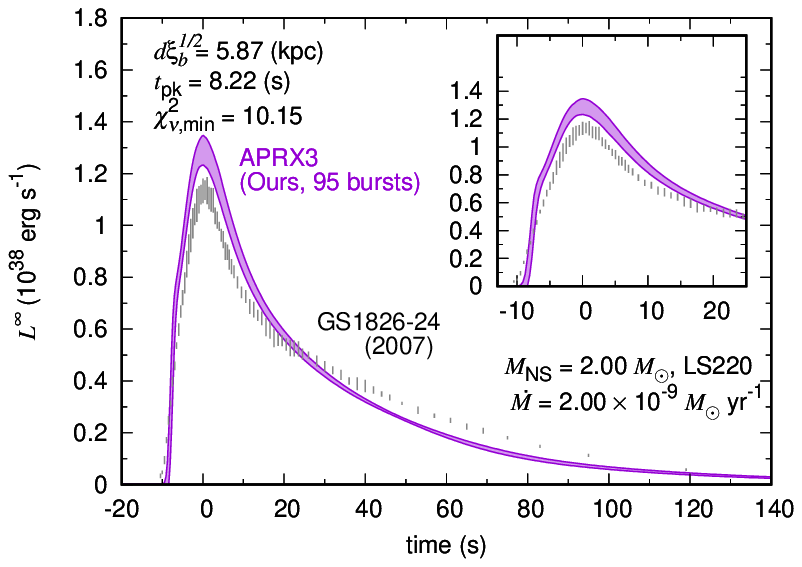}
\end{minipage}
\caption{Light-curves with $\dot{M}_{-9} = 2.0$ and $Z_{\rm CNO} = 0.01$. Left panel: {\it Best fit} model of $M = 1.58~M_{\odot}$. Right-panel: Model of  $M = 2.00~M_{\odot}$ whose overall shape is inconsistent with observations.}
\label{fig:bestfit}
\end{figure}

We have tried to reproduce the light curve observation of  X-ray bursts from GS 1826-24.
To properly calculate the X-ray bursts, it is needed to include  physical quantities such as EoS, nuclear energy generation rates, opacities,  and neutrino
loss rates  inherent to accreting neutron stars. Zamfir et al.~\cite{Zam2012} constrained neutron star mass and radius to be
$R<9.0-13.2$~km and $M<1.2-1.7~M_{\odot}$, respectively. However, they used a theoretical light curve model in Ref.~\cite{Heger2007a}
who assumed the radius $R=11.2$~km of $M=1.4~M_{\odot}$ and the gravitational redshift is $z=0.26$. 
In this paper, we focus on specific EoS which determines the mass and radius. As a result, by using the statistical method of
$\chi^2$, we reproduce the observational light curve with the standard deviation of $1\sigma$ level.
If we adopt another EoS, we may get different results. To see the dependence on EoS, we have shown the case of $K=180$~MeV in Table~\ref{tab:result}
, where the peak luminosity is rather low and the recurrence time becomes shorter. The incompressibility $K$ is the most uncertain quantity in EoS and
the softer one seems to be inconsistent with observations. It should be continued to study the effects of EoS properties on X-ray bursts. 
Concerning the burster GS 1826-24, the accretion rate is inferred to be around $10^{-9}~M_\odot~\rm yr^{-1}$ which is a rather 
high accretion rate to satisfy the condition of regular bursts. In this case, X-ray bursts will depend on the accretion layers and the
connection between the accretion layers and the stellar core is weak~\cite{Fujimoto1984}. 
By comparing the theoretical and observational light curves,  some physical processes such as a strong cooling
process should be examined based on theories and experiments of nuclear physics~\cite{Matsuo2018}. 

From Fig.~\ref{z005to2}, the shape of the theoretical light curve depends on $Z_{\rm CNO}$. 
If we chose  $\dot{M}_{-9} = 2.0$, $Z_{\rm CNO} = 0.01$ and $M = 1.58~M_{\odot}$, the tail of the light curve agrees
rather well compared to the observation as seen in Fig.~\ref{fig:bestfit}.
This is due to the energy generation of the $rp$-process. On the other hand,
many reaction rates concerning the nucleosynthesis will be responsible for the shape.  The uncertainty of the reaction rates has been guessed in the range of factor 100,  because of the poor understanding of the reaction rates in the proton-rich 
nuclei~\cite{Woosley2004,Meisel2018}. 
As a consequence, it is difficult to select key reactions important for the construction of the light curve~\cite{Meisel2019}.

Finally, we consider the ratio of the burst anisotropy  $\xi_{\rm b}$ to the persistent one  $\xi_{\rm p}$ between bursts. 
The anisotropy related to the persistent luminosity is obtained from $\dot{M}$ and the gravitational redshift $z_g$:
 $4\pi d^2\xi_{\rm p} F_x = \dot{M}c^2 z_g/(1+z_g)$ with the speed of light $c$~\cite{Heger2007a}.
Our best fit model L2n20Z1 gives $z_g = 0.26$, $\dot{M}=2\times10^{-9}~M_\odot~\rm yr^{-1}$, and the persistent flux
$F_x= 5.4\times10^{-9}~\rm erg~cm^{-2}s^{-1}$.
With the use of the peak flux and luminosity, $d^2\xi{\rm_b}$ is calculated  from 
$F_{\rm peak}=L_{\rm peak}/4\pi d^2\xi_{\rm b}$. As the result, the ratio $\xi_{\rm b}/ \xi_{\rm p}=1.01$.  
Furthermore, the inclination angle relative to the observer
becomes  $\theta\approx 58.6^\circ$ on the assumption of a thin flat disk~\cite{Fujimoto1988}.  It has been reported that while the angle becomes
$\theta\approx 80^\circ$ using the MESA code~\cite{Meisel2018}, the KEPLER best fit requires $\theta\approx 65^\circ$~\cite{Heger2007a}. 
Although it is complicated to constrain the disk model, it may be worthwhile to study the dependence of the anisotropy on EoS. 

We remark future studies concerning type I X-ray bursts which should be carried out based on the present work.

1. Detailed constraints for the equation of state beyond the nuclear saturation density will become possible by comparing
light curves between theories and observations, which include the uncertainty of  disk models.  

2. Whole reaction rates related to the $rp$-process should be restudied  from the point of nuclear experiments and theories.
  
3. Bursts with strong cooling coupled to parameters associated with the superfluidity model would be examined for lower accretion
rates $dM/dt \le 10^{-9}~M_\odot~\rm yr^{-1}$ and/or long intervals between bursts.

4. Long term simulations of X-ray bursts beyond one billion years could be desirable by using the approximate network 
since the beginning of bursts after the formation of low mass X-ray binary cannot be identified.  


\section*{Acknowledgment}
We thank T. Noda, and K. Arai for helpful discussions and are grateful for the referee's comments.
This work was supported by JSPS KAKENHI Grant Numbers 24540278 and 15K05083.


\small
\begin{thebibliography}{99}

\bibitem{Grindlay1976}
J.~Grindlay {\em et~al.},
\newblock \apj {\bf 205}, 127 (1976).

\bibitem{Monash2019}
https://burst.sci.monash.edu/wiki/index.php?n=MINBAR.SourceTable
\newblock .

\bibitem{Taam1982}
R.~E. {Taam},
\newblock \apj {\bf 258}, 761 (1982).

\bibitem{Wallace1981}
R.~K. Wallace and S.~E. Woosley,
\newblock \apj Suppl. {\bf 45}, 389 (1981).

\bibitem{Tanaka1989}
Y.~{Tanaka},
\newblock {Black holes in X-ray binaries: X-ray properties of the galactic
black hole candidates},
\newblock in {\em Two Topics in X-Ray Astronomy}, edited by J.~{Hunt} and
B.~{Battrick}, ESA Vol. 296, (1989).

\bibitem{Hashi2014}
M.~{Hashimoto} {\it et~al.},
\newblock J. Astrophys. {\bf 2014}, ID817986 (2014).

\bibitem{Cornelisse2003}
R.~{Cornelisse} {\em et~al.},
\newblock Astron. Astrophys. {\bf 405}, 1033 (2003).

\bibitem{Bildsten2000}
L.~{Bildsten},
\newblock {Theory and observations of Type I X-Ray bursts from neutron stars}, edited by S.~S. {Holt} and W.~W. {Zhang}, American Institute of Phys. Conf.  Vol. 522, pp. 359--369, (2000).

\bibitem{Fuji1981}
M.~Y. Fujimoto, T.~Hanawa, and S.~Miyaji,
\newblock \apj {\bf 246}, 269 (1981).

\bibitem{Heger2007a}
A.~Heger, A.~Cumming, D.~K. Galloway, and S.~E. Woosley,
\newblock \apj Lett. {\bf 671}, L141 (2007).

\bibitem{Lampe2016}
N.~Lampe, A.~Heger, and D.~K. Galloway,
\newblock \apj {\bf 819}, 1 (2016).

\bibitem{Galloway2017}
D.~K. Galloway,  A. J. Goodwin, and L. Keek,
\newblock PASA {\bf 34}, e019 (2017).

\bibitem{Meisel2018}
Z.~{Meisel},
\newblock \apj {\bf 860}, 147 (2018).

\bibitem{Meisel2019}
Z.~{Meisel},
\newblock \apj {\bf 872}, 84 (2019).

\bibitem{Woosley2004}
S.~E. Woosley {\em et~al.},
\newblock \apj Suppl. {\bf 151}, 75 (2004).

\bibitem{Zam2012}
M.~{Zamfir}, A. Cumming, and D. K. Galloway,
\newblock \apj {\bf 749}, 69 (2012).

\bibitem{Fujimoto1984}
M.~Y. Fujimoto, T.~Hanawa, J.~I. Iben, and M.~B. Richardson,
\newblock \apj {\bf 278}, 813 (1984).

\bibitem{Nara2003}
R. Narayan and J. S. Hey,
\newblock \apj {\bf 599}, 419 (2003).
\bibitem{Nara2006}
R. Cooper and R. Narayan,
\newblock \apj {\bf 652}, 584 (2006).

\bibitem{Dohi2019}
A. Dohi, K. Nakazato, M. Hashimoto, Y. Matsuo, and T. Noda,
Prog. Exp. Theor. Phys, {\bf 2019}, 113E01 (2019).
\bibitem{Lim2017}
Y. Lim, C. H. Hyun, and C.-H. Lee,
 Int. J. Mod. Phys. E, {\bf 26}, 1750015-328 (2017).
\bibitem{Oertel2017}
M. Oertel, M. Hempel, T. Kl$\ddot{\rm a}$hn, and S. Typel,
Rev. Mod. Phys. {\bf 89}, 015007 (2017).


\bibitem{Thorne1977}
K.~S. Thorne,
\newblock \apj {\bf 212}, 825 (1977).

\bibitem{Lattimer1991}
J.~M. Lattimer and F.~D. Swesty,
\newblock Nucl. Phys. A {\bf 535}, 331 (1991).

\bibitem{Hebe2013}
K. Hebeler, J. M. Lattimer, C. J. Pethick, and A. Schwenk, Astrophys. J. 773, 11 (2013).
\bibitem{Garg2018}
U. Garg, and G. Colo, Prog. Part. Nucl. Phys. 101, 55 (2018).

\bibitem{Matsuo2018}
Y. Matsuo, H. Liu, M. Hashimoto, and T. Noda,  Int. J. Mod. Phys. E,
{\bf 27}, 1850067 (2018).

\bibitem{Baym1971}
G.~Baym, C.~Pethick, and P.~Sutherland,
\newblock \apj {\bf 170}, 299 (1971).

\bibitem{Festa1969}
G.~G. {Festa} and M.~A. {Ruderman},
\newblock Phys. Rev. {\bf 180}, 1227 (1969).

\bibitem{Friman1979}
B.~L. {Friman} and O.~V. {Maxwell},
\newblock \apj {\bf 232}, 541 (1979).

\bibitem{Chiu1964}
H.-Y. {Chiu} and E.~E. {Salpeter},
\newblock Phys. Rev. Lett. {\bf 12}, 413 (1964).

\bibitem{Yin2017}
P.~{Yin}, X.~{Fan}, J.~{Dong}, W.~{Guo}, and W.~{Zuo}, Nucl. Phys. A
{\bfseries 961} 200 (2017).

\bibitem{Hanawa1983}
T.~{Hanawa}, D.~{Sugimoto}, and M.~{Hashimoto},
\newblock Publ. Astron. Soc. Jpn. {\bf 35}, 491
(1983).

\bibitem{Koike1999}
O.~{Koike}, M.~{Hashimoto}, K.~{Arai}, and S.~{Wanajo},
\newblock \apj {\bf 342}, 464 (1999).

\bibitem{Cyburt2010}
R.~H. {Cyburt} {\em et~al.},
\newblock \apjs {\bf 189}, 240 (2010).

\bibitem{Lam2016}
Y.~H. {Lam} {\em et~al.},
\newblock \apj {\bf 818}, 78 (2016).

\bibitem{Hashimoto1985}
M.~Hashimoto and K.~Arai,
\newblock Phys. Rep. of Kumamoto Univ. {\bf 7}, 47 (1985).

\bibitem{Oda1994}
T.~{Oda}, M.~{Hino}, K.~{Muto}, M.~{Takahara}, and K.~{Sato},
\newblock At. Data Nucl. Data Tables {\bf 56}, 231 (1994).

\bibitem{Langanke2000}
K.~{Langanke} and G.~{Mart{\'{\i}}nez-Pinedo},
\newblock Nucl. Phys. A {\bf 673}, 481 (2000).

\bibitem{Nabi1999}
J.-U. {Nabi} and H.~V. {Klapdor-Kleingrothaus},
\newblock At. Data Nucl. Data Tables {\bf 71}, 149 (1999); {\it ibid.},  {\bf 88}, 237 (2004).

\bibitem{Schatz2006}
H.~Schatz,
\newblock Int. J. Mass Spectrom. {\bf 251}, 293 (2006).

\bibitem{Matsuo2017}
Y.~{Matsuo}, Doctor thesis in Kyushu Univ. https://doi.org/10.15017/1806813
\newblock (2017).

\bibitem{Fujimoto1988}
M.~Y. Fujimoto,
\newblock \apj {\bf 324}, 995 (1988).



\end{thebibliography}

\end{document}